\documentstyle[preprint,eqsecnum,prd,aps]{revtex}

\begin{document}
\preprint{
\vbox{\halign{&##\hfil\cr
		& NUHEP-TH-96-5       	\cr}}}		
\input{epsf}
\title{ COMPARATIVE STUDY OF THE  
HADRONIC PRODUCTION OF $B_c$ MESONS\footnote{Presented at the American 
Physical Sociaty, Division of Particles and Fields Meeting,
10-15 August 1996.} }
\author{ ROBERT J. OAKES }
\address{Department of Physics and Astronomy, \\
Northwestern University, Evanston, IL 60208, USA }
\maketitle
\abstract{
The full order $\alpha_s^4$ perturbative QCD calculation of the production 
of $B_c$ mesons at the Tevatron is
compared with the fragmentation approximation. The non-fragmentation
diagrams, in which two or more quarks and/or gluons can simultaneously
be nearly on-shell, are important unless $P_T\gg M_{B_c}$.}

\section{Introduction}
Hadronic production of $B_c$ and $B_c^*$ mesons is of considerable, 
current interest.
At hadronic colliders, the dominant subprocess is $g + g \to B_c(B_c^*)
+ b + \bar{c} $ and 36 Feynman diagrams contribute in the lowest order,
$\alpha_s^4 $. At sufficiently large $B_c$ or $B_c^*$ transverse momentum,
$P_T$, the fragmentation approximation dominates. This process is ideal 
for quantitatively comparing the fragmentation approximation and the full
order $\alpha_s^4$ calculation, since both can be reliably calculated.
We have shown~\cite{cco} that the fragmentation approximation and the full
order $\alpha_s^4$ calculation agree when and only when $P_T \gg M_{B_c}$.
At small $P_T$ the non-fragmentation contributions become important since,
when the $B_c$ or $B_c^*$ is nearly collinear with the initial partons, it is
possible for two or more quarks and/or gluons in the subprocesses
to simultaneously be nearly on-mass-shell. 

\section{Calculations and Results}
Figure~1 shows the $P_T$ 
distributions for the process $p+\bar{p} \to B_c (B_c^*) +X $ at the Fermilab
Tevatron energy $\sqrt{s} = 1.8 $ TeV with the rapidity cut $|Y| < 1.5 $.
Remarkably, the $B_c$
meson $P_T$ distributions agree rather well, even for $P_T$ as small as 
about $5$ GeV, while for the $B_c^*$ meson the distributions differ by 
$50 - 70 \% $ over a much larger range of $P_T$, leaving the comparison
inconclusive.

The $P_T$ distributions for the subprocess $ g (k_1) + g (k_2) \to B_c 
(P) + b (q_2) +\bar{c} (q_1) $, which are shown in Fig.~2, are somewhat
more revealing. The distributions agree reasonable well when $P_T$ is
larger than about $30$ GeV for the $B_c$ meson and about $40$ GeV 
for the $B_c^*$ meson. Consequently, the $P_T$ distributions alone 
are not decisive and, as will be shown, can even be misleading.

It is more insightful to examine the distributions in 
$z = 2 (k_1 + k_2 ) \cdot P /\hat{s} $, which  is simply twice the fraction
of the total energy carried by the $B_c$ or $B_c^*$ in the subprocess center
of mass. Note that $z$ is experimentally measurable, at least in principle. 
Figure~3 shows the $z$ distributions, $C(z)$, for the process 
$p+\bar{p} \to B_c (B_c^*) +X $ for $ \sqrt{s} = 1.8 $ TeV, $ |Y| < 1.5 $,
and $ P_T > 10$ GeV. The distribution~\cite{cco} $C(z)\equiv d\sigma / dz$. 
From Fig.~3 it is clear that for the $B_c$ the 
fragmentation approximation 
underestimates the full order $\alpha_s^4$ calculation for small $z$ and 
overestimates it for large $z$. This results in a cancelation in the $B_c$
$P_T$ distribution, Fig.~1, fortuitously causing the fragmentation
approximation to agree with the full order $\alpha_s^4$ calculation down
to quite small values of $P_T$. However, for the $B_c^*$ the fragmentation
calculation underestimates the full order $\alpha_s^4$ calculation at all
values of $z$ and no fortuitous cancelation occurs, as for the $B_c$.

The relative importance of the fragmentation and non-fragmentation 
contributions is even more clearly evident in the $P_T$ distributions for the
subprocess 
$g + g \to B_c (B_c^*) + b + \bar{c} $ at very large $ \sqrt{\hat{s}} $, 
as is shown in Fig.~4.
Clearly, the non-fragmentation contributions dominate
at small $P_T$, where two or more quarks and/or gluons can 
simultaneously be very nearly on-mass-shell in the subprocess.

\section{Conclusions} 
 We have compared the full order $\alpha_s^4$ perturbative QCD 
calculation of the production of $B_c$ and $B_c^*$ mesons
at the Fermilab Tevatron with the fragmentation approximation. 
There are Feynman diagrams present in the full order $\alpha_s^4$
matrix element in which {\it two or more } quarks and/or gluons
can simultaneously be nearly on-mass-shell, and these dominate over the 
fragmentation approximation at small $P_T$. The fragmentation
approximation dominates when and only when $P_T \gg M_{B_c}$.

\section*{Acknowledgments} 
This research was done in collaboration with Chao-Hsi Chang and 
Yu-Qi Chen and is presented in greater detail in reference 1, where
references to previous work can also be found. We acknowledge the 
support of the U.S. Department of Energy, Division of High Energy
Physics, under Grant DE-FG02-91-ER40684 and the National Nature
Science Foundation of China.

\begin{figure}[h] 
\begin{center}
\leavevmode
\epsfysize=120mm
\epsfbox{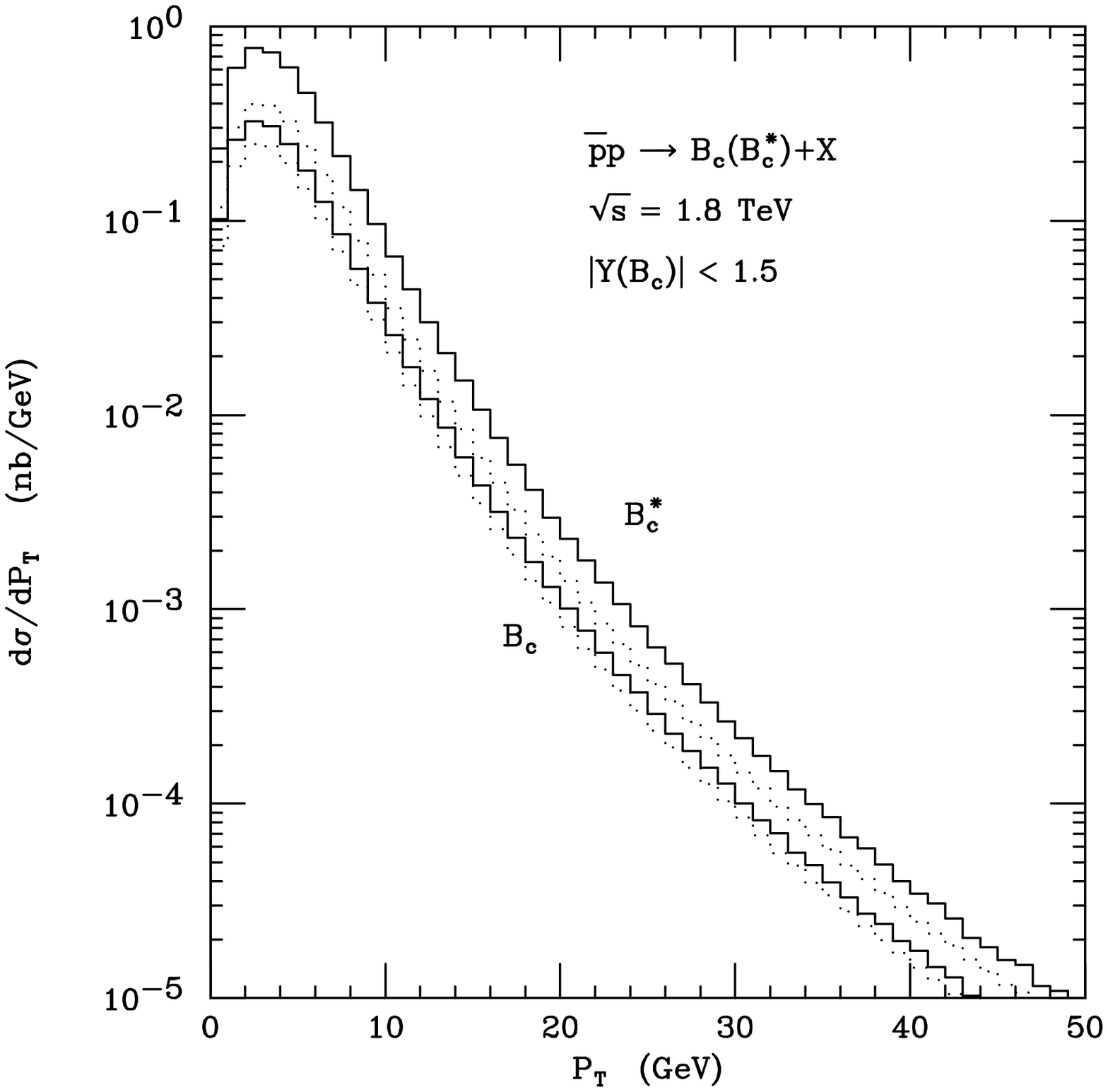}
\end{center}
\footnotesize
Fig. 1~  The $P_T$ distributions for $B_c$ and $B_c^*$ meson 
production at the Tevatron energy $\sqrt{{s}}=$ 1.8 TeV. The solid  
and the doted lines correspond to the full $\alpha_s^4$ calculation 
and the fragmentation approximation, respectively. 
\end{figure}

\begin{figure}[h] 
\begin{center}
\leavevmode
\epsfysize=120mm
\epsfbox{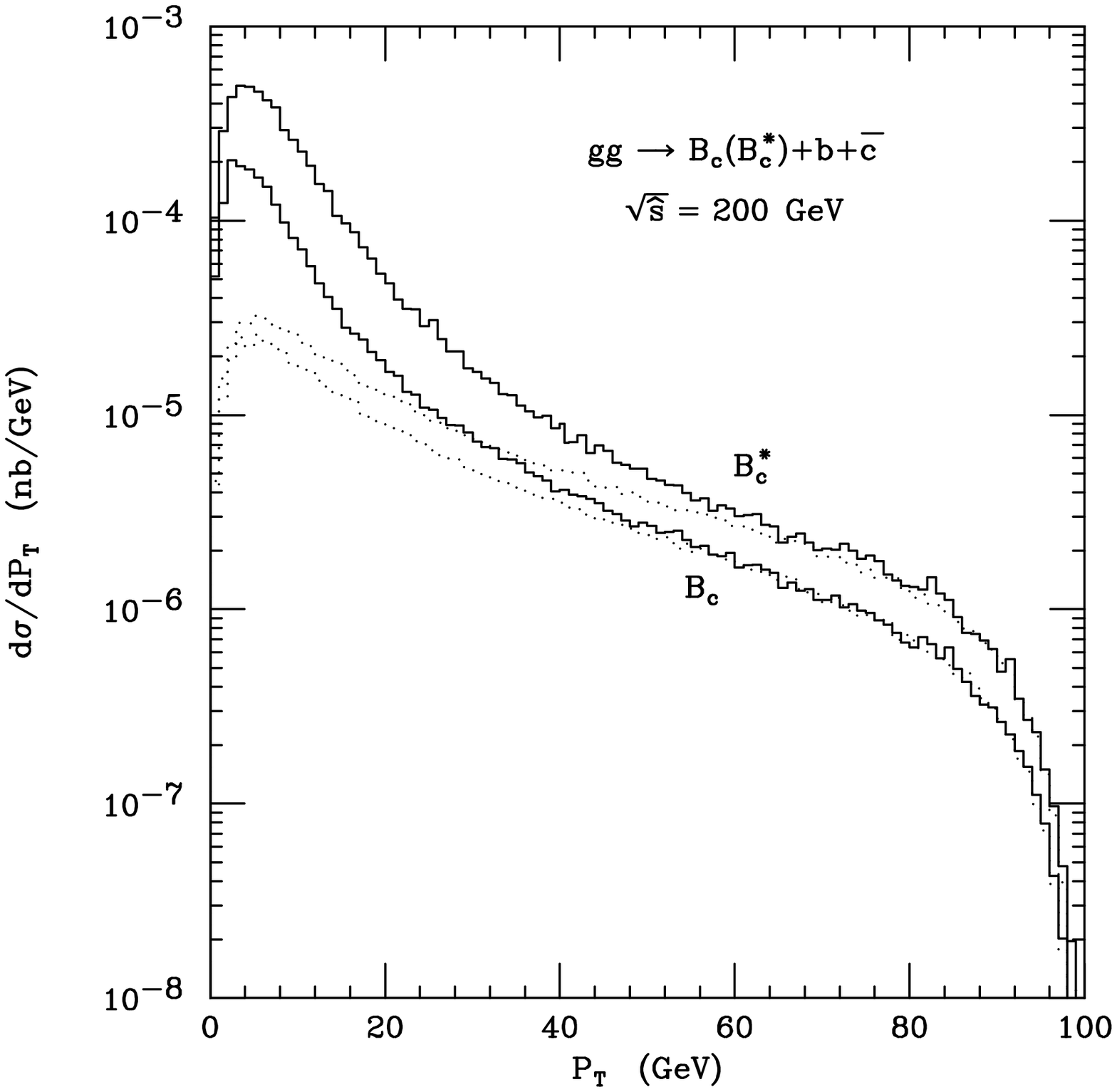}
\end{center}
\footnotesize
Fig. 2~ The $P_T$ distributions of the $B_c$ and $B_c^*$ meson for 
the subprocess with $\sqrt{\hat{s}}=$ 200 GeV. The solid and the doted 
lines correspond to the full $\alpha_s^4$ calculation and the fragmentation 
approximation, respectively.
\end{figure}

\begin{figure}[h] 
\begin{center}
\leavevmode
\epsfysize=120mm
\epsfbox{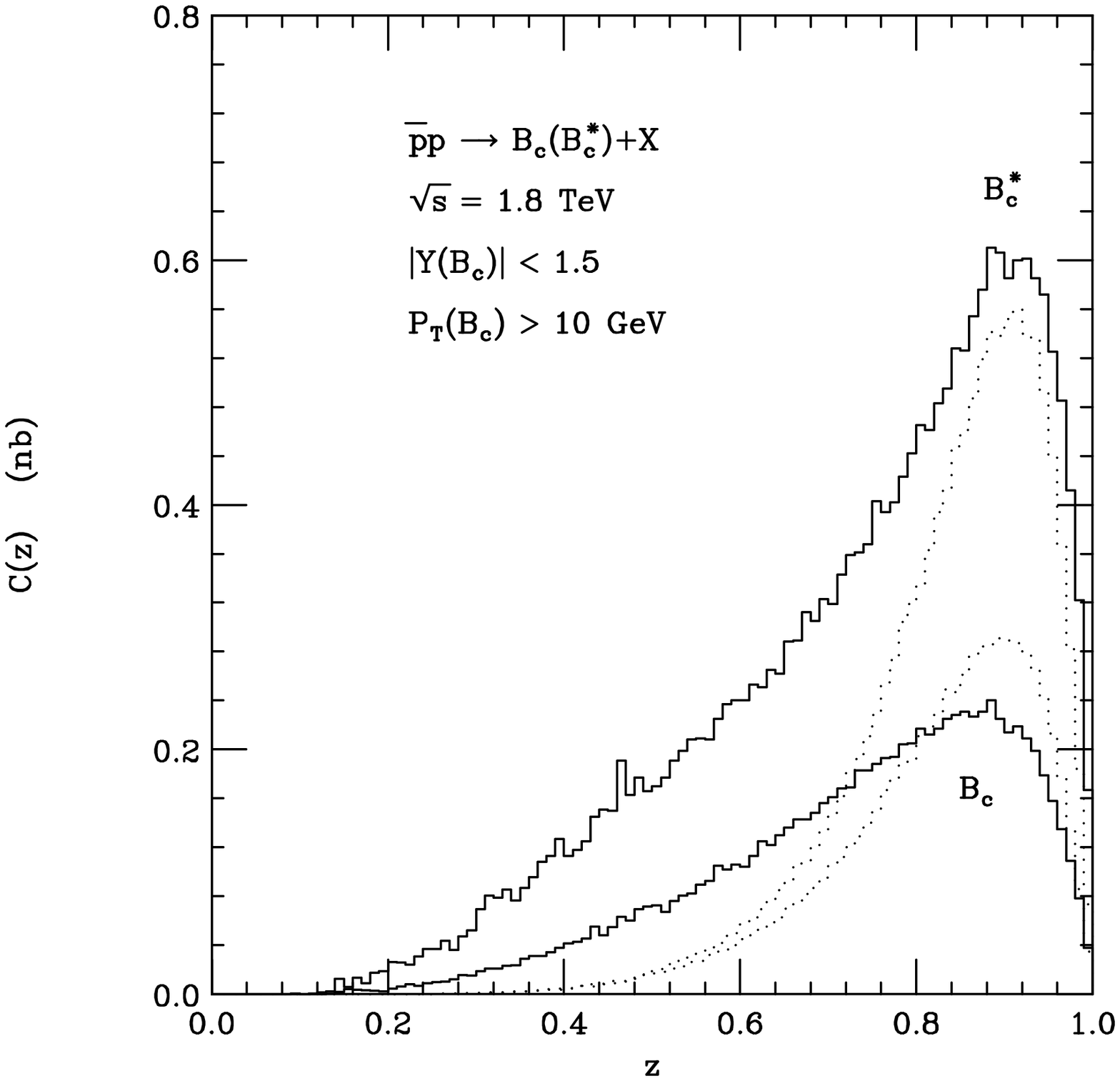}
\end{center}
\footnotesize
Fig. 3~ The $z$ distributions $C(z)$ of the $B_c$ and $B_c^*$ 
at the Tevatron energy $\sqrt{s}=1.8$ TeV. The solid lines are the
full $\alpha_s^4$ calculation and the doted lines are the fragmentation
approximation with the cut $P_T> 10$ GeV  
\end{figure}

\begin{figure}[h] 
\begin{center}
\leavevmode
\epsfysize=120mm
\epsfbox{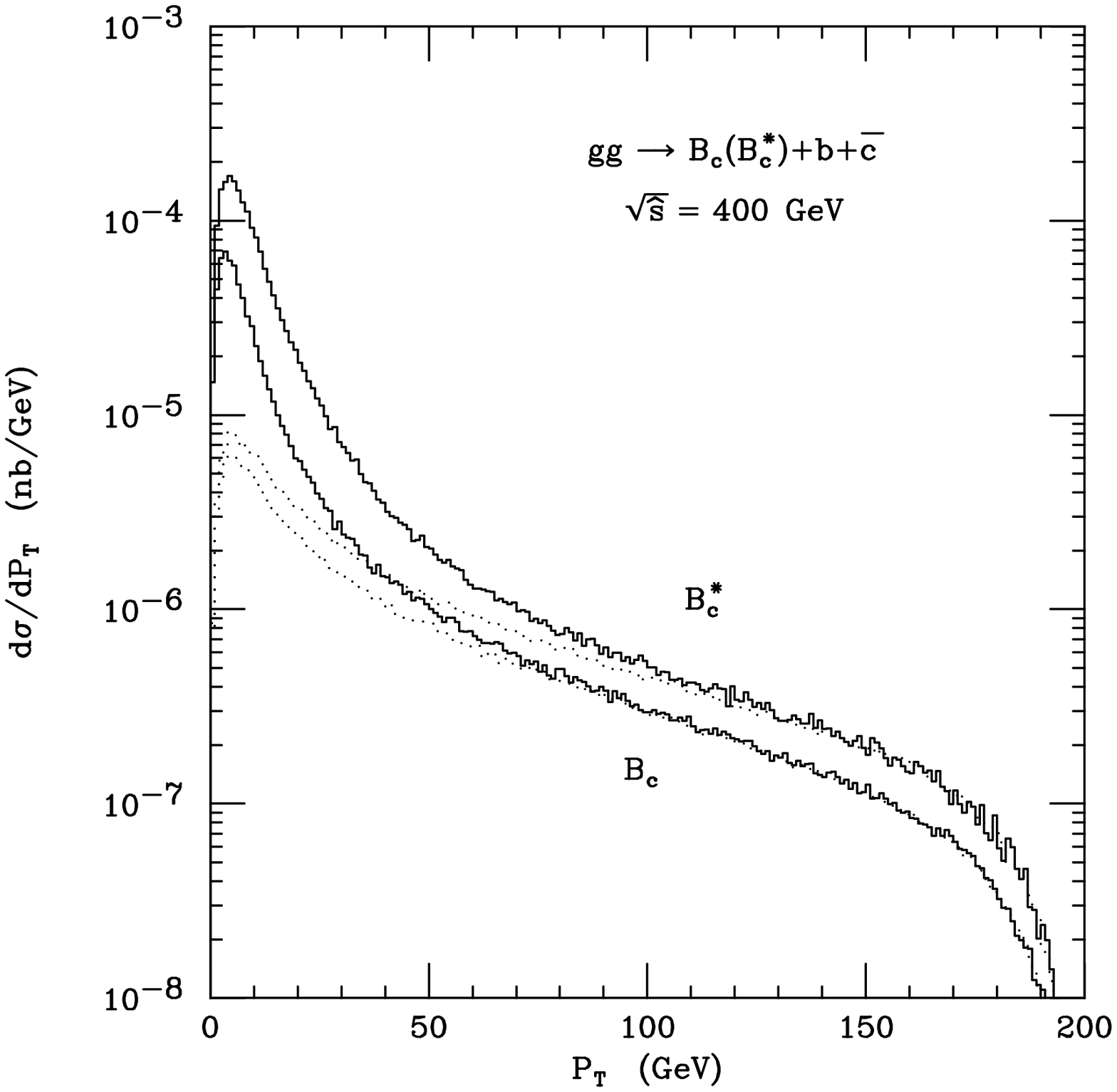}
\end{center}
\footnotesize
Fig. 4~ The $P_T$ distributions of the $B_c$ and $B_c^*$ meson for 
the subprocess with $\sqrt{\hat{s}}=$ 400 GeV. The solid and the doted 
lines correspond to the full $\alpha_s^4$ calculation and the fragmentation 
approximation, respectively.
\end{figure}

\end{document}